\documentclass[useAMS,usenatbib]{mn2e}
\usepackage[dvips]{graphicx}

\newcommand{\ee}{$e^\pm$}

\newcommand{\lh}{\ell_{\rm h}}
\newcommand{\ls}{\ell_{\rm s}}
\newcommand{\lth}{\ell_{\rm th}}
\newcommand{\lnth}{\ell_{\rm nth}}
\newcommand{\msun}{{\rm M}_{\sun}}
\newcommand{\ledd}{L_{\rm E}}
\newcommand{\xte}{{\it RXTE}}

\newcommand{\asm}{{\it RXTE}/ASM}
\newcommand {\batse}{{\it CGRO}/BATSE}
\newcommand {\osse}{{\it CGRO}/OSSE}
\newcommand{\integral}{{\it INTEGRAL}}

\title[The nature of the hard state of Cygnus X-3]{The nature of the hard state of Cygnus X-3} 

\author[L. Hjalmarsdotter et al.]
{L.~Hjalmarsdotter$^{1,2}$\thanks{E-mail:nea@astro.su.se}, 
A. A.~Zdziarski$^{3}$, S.~Larsson$^{2}$, V.~Beckmann$^{5,6}$,  
\newauthor M.~McCollough$^{7}$, D.~C.~Hannikainen$^{1}$, O.~Vilhu$^{1}$\\
$^{1}$Observatory, PO Box 14, FIN-00014 University of Helsinki, Finland\\
$^{2}$Stockholm Observatory, Department of Astronomy, AlbaNova University Center, 106 91 Stockholm, Sweden\\
$^{3}$Copernicus Astronomical Center, Bartycka 18, 00-716 Warszawa, Poland\\
$^{5}$INTEGRAL Science Data Centre, Ch. d'\'Ecogia 16, 1290 Versoix, Switzerland\\
$^{6}$Joint Center for Astrophysics, Department of Physics, University of Maryland, Baltimore County, MD 21250, USA\\
$^{7}$Smithsonian Astrophysical Observatory, 60 Garden Street, MS 67, Cambridge, MA 02138-1516, USA}

\begin{document}


\pagerange{\pageref{firstpage}--\pageref{lastpage}}
\pubyear{2007}

\maketitle

\label{firstpage}

\begin{abstract} 
The X-ray binary Cygnus~X-3 is a highly variable X-ray source that displays a wide range of observed spectral states. One of the main states is significantly harder than the others, peaking at $\sim20$ keV, with only a weak low-energy component. Due to the enigmatic nature of this object, hidden inside the strong stellar wind of its Wolf-Rayet companion, it has remained unclear whether this state represents an intrinsic hard state, with truncation of the inner disc, or whether it is just a result of increased local absorption. We study the X-ray light curves from \asm\/ and \batse\/ in terms of distributions and correlations of flux and hardness and find several signs of a bimodal behaviour of the accretion flow that are not likely to be the result of increased absorption in a surrounding medium. 
Using \integral\/ observations, we model the broad-band spectrum of Cyg~X-3 in its apparent hard state. We find that it can be well described by a model of a hard state with a truncated disc, despite the low cut-off energy, if the accreted power is supplied to the electrons in the inner flow in the form of acceleration rather than thermal heating, resulting in a hybrid electron distribution and a spectrum with a significant contribution from non-thermal Comptonization, usually observed only in soft states. The high luminosity of this non-thermal hard state implies that either the transition takes place at significantly higher $L/\ledd$ than in the usual advection models, or the mass of the compact object is  $\ga 20 \msun$, possibly making it the most massive black hole observed in an X-ray binary in our Galaxy so far.
We find that an absorption model as well as a model of almost pure Compton reflection also fit the data well, but both have difficulties explaining other results, in particular the radio/X-ray correlation. 
\end{abstract}


\section{Introduction to Cyg~X-3}
Cygnus X-3 is one of the brightest X-ray binaries, discovered already 40 years ago (Giacconi et al. 1967), extensively studied, but still poorly understood and usually referred to as a `peculiar source'. The identification of the donor as a Wolf-Rayet star (van Keerkwijk et al. 1992) classifies it as a high mass X-ray binary, despite its typical low mass binary period of 4.8 hours (Parsignault et al. 1972). It shows signs of unusually strong and complex absorption, a consequence of the whole system being enshrouded in the wind of its companion Wolf-Rayet star. The problem in separating wind features from those arising in the photosphere of the companion makes determinations of radial velocity difficult, hence the masses of the components remain uncertain and the nature of the compact object unknown. Published results range from a neutron star of $1.4\msun$ (Stark \& Saia 2003) to a $17\msun$ black hole (Schmutz, Geballe \& Schild 1996). 
Adding to its peculiarity, Cyg~X-3 is also the strongest radio source associated with an X-ray binary, displaying huge flares and relativistic jets (Mart{\'{\i}} et al. 2000, Mioduszewski et al. 2001). The system is located at a distance of 9 kpc (Dickey 1983, assuming 8 kpc for the distance to the galactic centre; Predehl et al. 2000), close to the Galactic plane.

\begin{center}
\begin{figure}
\includegraphics[angle=0,width=8cm, height=8cm]{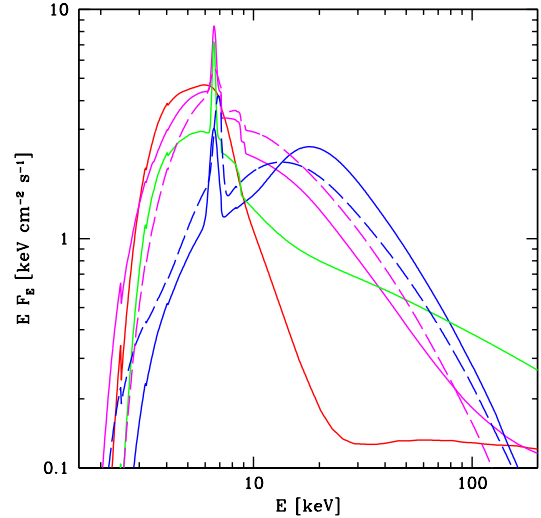} 
\caption{Example average spectral shapes of Cyg X-3, not corrected for absorption, as observed by \xte\/ (solid lines, Szostek \& Zdziarski 2004), and \integral\/ (dashed lines, Hj04a). The \integral\/ spectra have here been re-normalized to the level of SPI, and therefore differ by a factor of two from those in the original publication.}
\label{groups}
\end{figure}
\end{center}

The study of the X-ray spectrum and its variability has suffered severely from the fact that we do not fully understand the properties of the surrounding medium. This has led to that a detailed interpretation of the Cyg~X-3 intrinsic unabsorbed spectrum and luminosity are still missing from the literature. The first physical interpretation of the Cyg~X-3 broad-band X-ray spectrum was presented in Vilhu et al. (2003), hereafter V03, based on simultaneous \integral\/ and \xte\/ observations in December 2002, when the source displayed intermediate to high flux-levels. The spectrum turned out to be well fitted with thermal Comptonization including Compton reflection and with parameters similar to other X-ray binaries at high accretion rates, with the addition of strong absorption. In a study of all available \xte\/ observations between 1996 and 2000, Szostek \& Zdziarski (2004) found that they could be divided into 5 states, with their absorbed appearances resembling the canonical states of X-ray binaries (see examples in Fig.~\ref{groups}). Hjalmarsdotter et al. (2004a), hereafter Hj04a, found that two of these states had also been observed by \integral\/ on several occasions, the one described in V03 and another state peaking at around 20 keV, resembling the canonical hard state, but with the cut-off at considerably lower energy. The observed {\it INTEGRAL} spectra could both be fitted using the same model as in V03, including very strong absorption. Unabsorbed spectral shapes were however not discussed in any of these publications and it was not clear whether the observed differences between states were due to an intrinsic state transition or just a result of varying absorption. 

The existence or absence of a transition into a hard state is crucial for the understanding of this system and its spectral variability. In this paper, we investigate the possibility of a state transition in Cyg~X-3, using results from a study of the X-ray light curve as well as broad band spectral modelling of recent \integral\/ data. Our aim is to determine whether the apparent hard state can be `real' in the sense that it represent an intrinsic transition into a non-disc dominated lower accretion rate state with truncation of the inner disc, despite the low, $\sim 20$ keV, cut-off energy, as compared to other sources. The alternative is that the apparent state transition is just an artefact of increased absorption.

\section{The X-ray light curve}
\subsection{Flux and hardness correlations}
\begin{figure}
\includegraphics[angle=0,width=8cm,height=6.5cm]{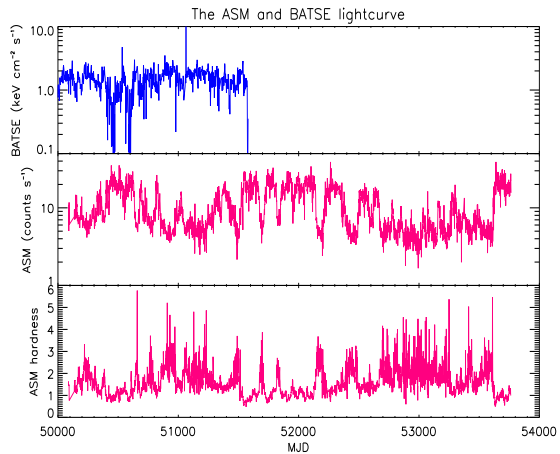}
\caption{{\it Top panel}: The \batse\/ 20--230 keV 3-day flux averages between 1996--2000. {\it Middle panel}: The \asm\/ 1.3-12 keV light daily count rate averages between 1996--2006. A count rate of 75 counts s$^(-1)$ corresponds to 1 Crab. {\it Lower panel}: The (5--12)/(1.3--5) keV hardness.}
\label{lc}
\end{figure}

Fig.~\ref{lc} shows 10 years of continuous monitoring of Cyg~X-3 in soft X-rays (1.3--12 keV) by the \asm\/ together with the 5 years of overlapping data in hard X-rays (20--230 keV) from \batse. The light curves are highly variable and show periods of high and low flux levels, anticorrelated with each other (McCollough et al. 1997). The (5--12)/(1.3--5) keV hardness in the lower panel is also anti-correlated with the 1.3--12 keV flux (Watanabe et al. 1994). Periods with high, soft ASM flux, and low to very low BATSE flux correspond to the range of observed soft states (red, magenta and green in Fig.~\ref{groups}), while the periods with low and hard ASM flux and higher BATSE flux correspond to the apparent hard state (blue in Fig.~\ref{groups}).

\begin{figure}
\includegraphics[width=0.44\textwidth]{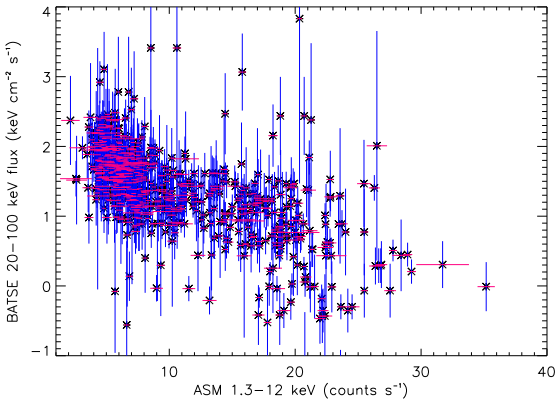}
\caption{The \batse\/ 20--100 keV flux plotted against the \asm\/ 1.3--12 keV count rate.}
\label{ffplot}
\end{figure}

\begin{figure}
\includegraphics[width=0.44\textwidth]{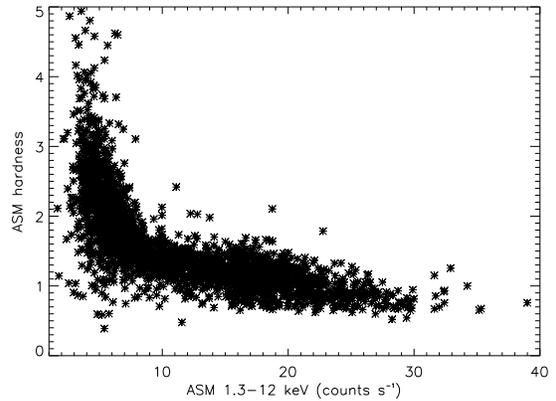} 
\caption{The (5--12)/(1.3--5) keV hardness plotted against the \asm\/ 1.3 --12 keV count rate.}
\label{hrplot}
\end{figure}

The anticorrelations between ASM and BATSE fluxes, also shown in Fig.~\ref{ffplot}, and between ASM flux and hardness, also in Fig.~\ref{hrplot}, are similar to those found in Cyg~X-1 in the hard state (Zdziarski et al 2002). In Cyg~X-3, these anticorrelations are present in all states, while their slopes, especially that of the flux-hardness anticorrelation in Fig.~\ref{hrplot}, changes abruptly at flux levels corresponding to the apparent state transition. 

\begin{figure*}
\includegraphics[width=\textwidth]{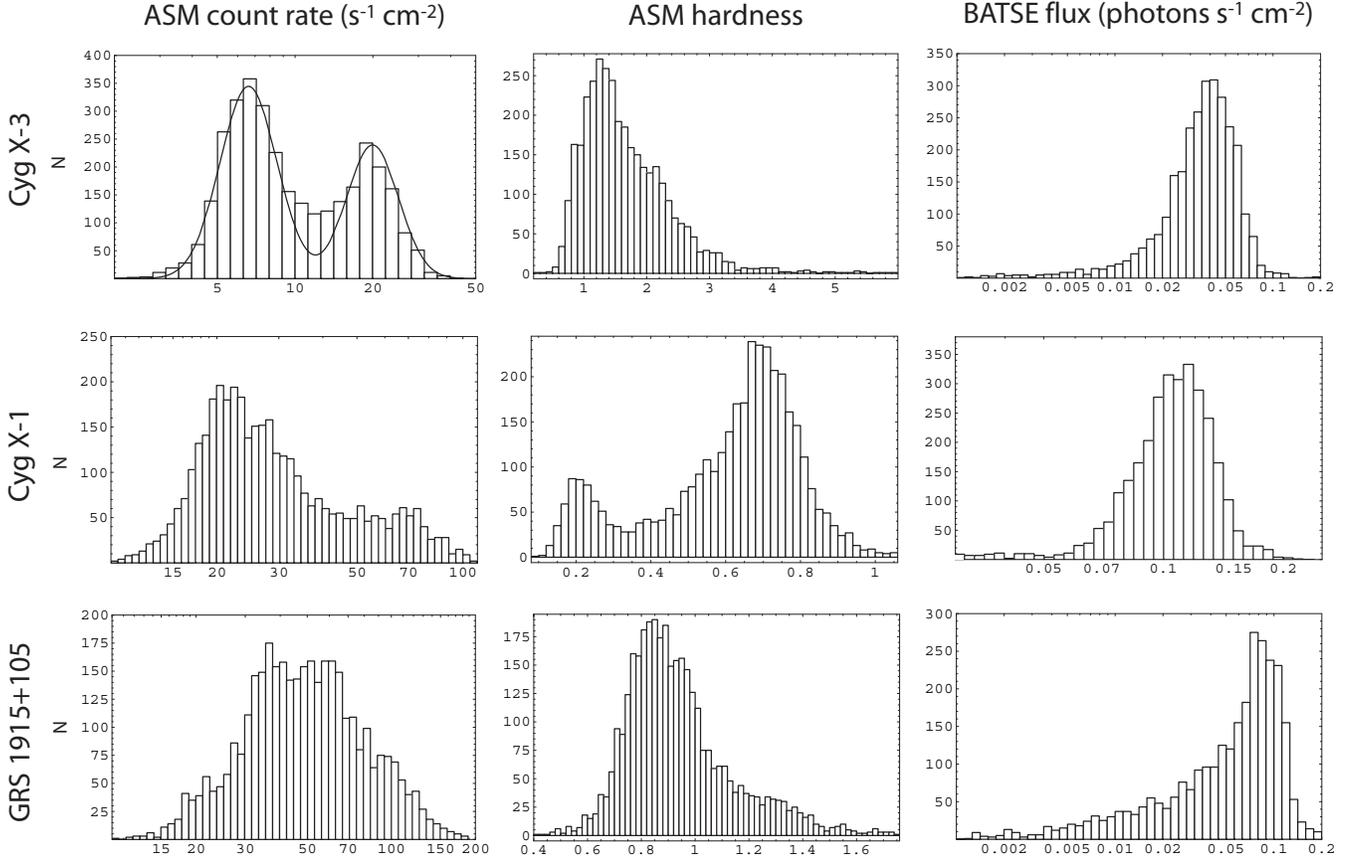}
\caption{Flux and hardness distributions of Cyg~X-3 (top), Cyg~X-1 (middle) and GRS~1915+105 (bottom). {\it Left column:} The distribution of count rates in the \xte/ASM 1.3--12 keV light curve, daily averages 1996--2006. {\it Middle column:} The distribution of the 5--12/1.3--5 keV hardness created from the \xte/ASM 1.3--12 keV light curve. {\it Right column:} The distribution of photon flux in the \batse\/ 20--100 keV light curve, daily averages 1991-2000.}
\label{hist}
\end{figure*}

\subsection{Distributions}
In Fig.~\ref{hist}, top left panel, the distribution of soft X-ray flux (1.3--12 keV) in Cyg~X-3 is plotted as a histogram. The distribution is bimodal, with the two peaks representing the soft and the apparent hard state respectively. The flux level varies more in the soft than in the hard state, but the distribution within each of the two states is well described as log-normal. The distribution of hardness in the top centre panel as well as that of the hard X-rays in the top right panel do not show the same bimodality, despite their overall anticorrelation with the soft X-ray flux. This is in agreement with the observed spectral shapes suggesting that the source exhibits a range of soft states, with a stable strong soft component, presumably from an accretion disc, and different amplitudes and slopes of the high energy tail (cf Fig.~\ref{groups}). The  existence of two well-defined intensity states in Cyg X-3 was in in fact suggested already by Watanabe et al. (1994), based on {\it Ginga} data, who suggested it a sign of a bimodal behaviour of an accretion disc.

It is interesting to make a comparison of the distributions of flux and hardness in Cyg~X-3 to those of other sources that do and to those that do not show state transitions into a hard state. The second row of Fig.~\ref{hist} shows the same distributions for Cyg~X-1. In this source, state transitions are less frequent than in Cyg~X-3, and it spends most time in the hard state. This fact, together with a variable flux level in its soft state, results in a different flux distribution from that of Cyg~X-3. A clear bimodality is however present in the distribution of hardness, showing the characteristic behaviour observed in this as well as in many other black hole binaries, in the transition between a soft, disc dominated state, and a hard, possibly advection dominated, state where the inner disc is truncated and only a weak disc component is observed. Just as in Cyg~X-3, the bimodality is not present in the distribution of hard X-ray flux, as observed by BATSE, indicating a more continuous variability pattern at higher energies.

The third row shows the same distributions for GRS~1915+105, the brightest Galactic X-ray binary, harbouring a $14 \msun$ black hole. In this source, the luminosity never falls below $0.3 \ledd$, which prevents it from entering a hard state (Done, Wardzi\'nski \& Gierli\'nski 2004). Its spectral variability is limited to that of varying strength of the high energy tail, with a strong disc component always present. This variability does not show a bimodal pattern in either the soft or hard flux nor in the hardness distribution.

\subsection{Orbital modulation}
Apart from the observed state transitions, the main variability pattern of the light curve at all energies is a strong modulation on the orbital period of 4.8 hours (Parsignault et al. 1972). The modulation is quasi-sinusoidal in shape and present in both soft and hard X-rays, with the shape and the place of the minima changing slightly with energy (V03, Hjalmarsdotter et al. 2004b, hereafter Hj04b). The preferred explanation for the modulation is attenuation in the wind of the companion (Willingale, King \& Pounds 1985; Kitamoto et al. 1987), even if obscuration by a disc bulge or accretion disc corona (White \& Holt 1982) still remains a possibility. 

\begin{figure}
\includegraphics[angle=0, width=8.5cm]{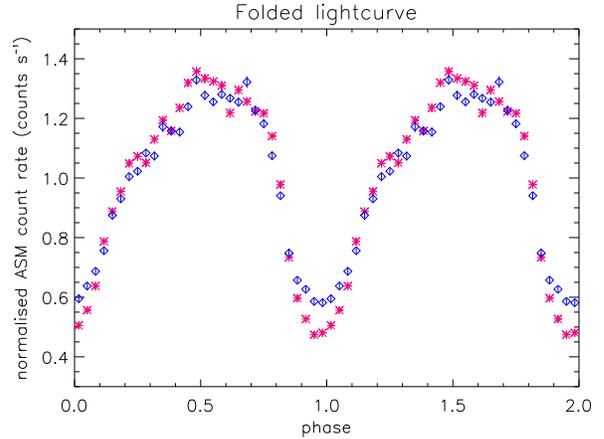}
\caption{The ASM 1.3--12 keV light curve from Fig.~\ref{lc} folded over the orbital period for the soft state(s) (ASM counts/s $>20$), magenta crosses, and the hard state (ASM counts/s $<10$), blue diamonds, respectively.}
\label{folds}
\end{figure}

If a dense stellar wind serves as the cause of the modulation, any change in its optical depth and/or distribution would be reflected in a change in the depth and/or shape of the modulated light curve, since the wind distribution is centered on the companion Wolf-Rayet star and thus asymmetric to the X-ray source. In particular, if the changes in absorption were large enough to cause the apparent state transition, they are also likely to affect the orbital modulation. An important aspect is therefore whether the strength of the modulation changes between the apparent hard and soft spectral states. To investigate this we selected sections of the ASM light curve, based on daily averaged ASM flux levels: $<10$ s$^{-1}$, representing the apparent hard state, and $>20$ s$^{-1}$, representing soft states, and folded them separately, using the parabolic ephemeris of Singh et al. (2002). The results are shown in Fig.~\ref{folds}. We find that the modulation is clearly present in both states, with a strength of approximately a factor of two in the hard state and a factor of three in the soft state(s). The shape of the modulation remains nearly constant. In fact, the difference in shape of the folded light curve between the two states is smaller than between different energy bands of the ASM data (when folding the entire light curve, see V03 and Hj04b), and is probably mostly due to the fact that the contribution of emission in the three ASM bands differs between the hard and soft states.

\subsection{Radio/X-ray correlation}
In recent years, a standard scheme for radio/X-ray correlations in black hole X-ray binaries seems to have been established. According to this scheme, steady radio emission, presumably from a compact jet, is present throughout the hard state, and is then positively correlated with the soft X-ray flux. In the soft state, the radio emission becomes strongly suppressed due to the disappearance of the compact jet. A transient jet episode sometimes manifests itself as a radio flare in the transition phase (eg. Fender, Belloni  \& Gallo 2004 and references therein).

\begin{figure}
\centering
\includegraphics[angle=0,width=8cm,height=7cm]{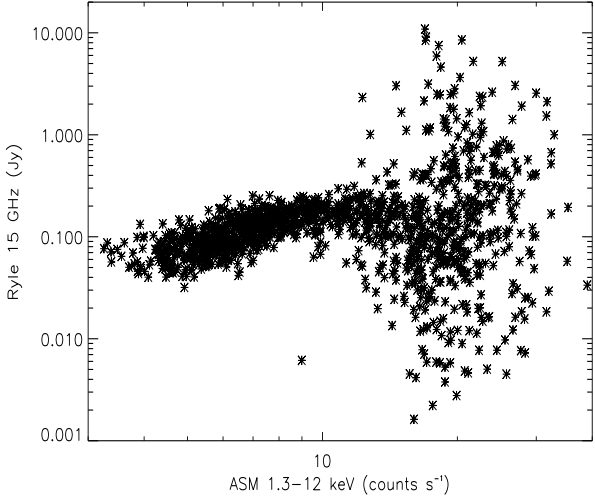} 
\caption{The 15 GHz flux density as observed by the Ryle telescope plotted against the ASM 1.3--12 keV count rate.}
\label{radio}
\end{figure}

Correlations between radio and X-ray emission in Cyg~X-3 have been reported by several authors (eg. Watanabe et al. 1994, McCollough et al. 1997, Choudhury  et al. 2002). Apart from the strong flaring episodes, observed at high soft X-ray flux levels in Cyg~X-3, and the very high radio to X-ray ratio, about an order of magnitude higher than in other sources, the overall correlation pattern in Cyg~X-3 is in fact similar to that of other black hole X-ray binaries (Gallo, Fender \& Pooley 2003).
Fig.~\ref{radio} shows the radio flux from Cyg~X-3, as measured at 15 GHz by the Ryle telescope, plotted against the ASM 1.3--12 keV soft X-ray count-rate. There is a clear difference in the radio behaviour on either side of the limit marking the apparent state transition. At low ASM flux levels, the radio emission is rather stable with flux densities ranging from $\sim 50 - 300$ mJy, and positively correlated with the ASM flux. At ASM flux levels above $\sim20$ s$^{-1}$, the radio behaviour is completely different, ranging from quenching at levels below 10 mJy, similar to that observed in other sources, to the giant radio flares reaching above 10 Jy. It is interesting to here once again compare the behaviour of Cyg~X-3 to that of Cyg~X-1 and GRS~1915+105, respectively. All three sources are included in the radio vs. \asm\/ flux plot in Gallo, Fender \& Pooley (2003). Cyg~X-1 displays radio behaviour in accordance with the scheme outlined in the beginning of this section. Throughout the hard state, the radio emission is stable and positively correlated with the ASM flux. At flux levels corresponding to the soft state, the radio emission is quenched. The same seems to be true for all Galactic black hole systems which experience a state transition between a hard state with a truncated disc and a soft disc dominated state. The strong radio flares displayed by Cyg~X-3 at high X-ray flux levels are, however, not observed. In GRS~1915+105, that never displays a hard state, the left-hand branch of stable radio emission is not present. Its radio behaviour is instead similar to that of Cyg~X-3 at high ASM flux levels, above the limit marking the state transition, with a large spread in amplitude between flares and quenching.

\section{Modelling the apparent hard state}
\subsection{Observations and data reductions}
To model the broad-band spectrum of the apparent hard state in Cyg~X-3 we selected an observation of the Cygnus region in November 2004, where Cyg~X-3 was in the field of view of all three X/$\gamma$-ray instruments on board \integral\/ for a period of nine days throughout revolutions 251, 252 and 253, 2--11 November, plus a few pointings from revolution 255, 14--17 November. During all this time, ASM flux levels were well below 10 s$^{-1}$. We only used pointings where Cyg~X-3 was within $3.5\degr$ off axis for JEM-X and within $10\degr$ off axis for IBIS/ISGRI and SPI, resulting in 64 pointings for JEM-X and 127 for IBIS/ISGRI and SPI respectively (total exposure time of 128 and 254 ksec). The JEM-X and ISGRI spectra were extracted using version 5.1 of ISDC's Off-line Scientific Analysis (OSA) software (Courvoisier et al. 2003). For JEM-X we used a detector radius limit of 108 mm, to avoid problems with increasing background close to the edges of the detector. Extractions were done at source positions determined from the imaging steps.
For each pointing, using the standard pipeline, we extracted the spectra 
of all the active sources in the field using a 256 channel response for JEM-X and a rebinned, 28 channels response matrix for IBIS/ISGRI. We then averaged the Cyg X-3 single pointing spectra into a final 
one for JEM-X, covering 3--25 keV, and for IBIS/ISGRI, covering 20--300 keV. The SPI data were analysed using image reconstruction and spectral extraction version 9.2 of the SPI Iterative Removal Of
Sources program (SPIROS; Skinner \& Connell 2003). We applied a background
model based on the mean count modulation of the
detector array. In order to get precise fluxes, the source positions of
the known sources in the field of view were fixed to their catalogue
values. In addition, we allowed SPIROS to apply time dependent
normalisation to the source fluxes of the other bright and variable
sources in the field of view (Cyg~X-1 and EXO 2030+375). 
For spectral extraction, 50 logarithmic bins in the 20--300 keV energy
range were used to create a SPI spectrum averaged over the total observation time. The instrumental response function has been derived
from on-the-ground calibration (Sturner et al. 2003) and then corrected
based on the Crab calibration observation.

The resulting broad-band spectrum, covering an energy range of 3--300 keV, is similar in shape to the hard state spectrum as observed by \integral\/ in Hj04a and by \xte\/ in Szostek \& Zdziarski (2004). However, while those spectra were both co-additions of data from several different observations, this one is the result of one continuous block of observations, and thus less affected by effects of averaging. Inspection of individual JEM-X and ISGRI spectra before co-addition show no variability between the different pointings other than in normalization as a result of the orbital modulation (and a slight increase in absorption around phases 0.2--0.4, in agreement with the difference in the shape of the folded light curves of different energies at those phases, V03, Hj04b). The spectrum was fitted in {\tt xspec} (Arnaud 1996), version 11.3. A systematic error of 2 percent was added to all three datasets. The JEM-X and ISGRI data were normalized to the SPI data, which has the most accurate absolute calibration according to the latest cross-calibration documentation, making the observed spectra agree well with the normalization of the \xte\/ spectra observed in similar states. By comparison, the spectra from V03 and Hj04a, that were normalized to JEM-X, were shown by V03 to be a factor of 2 too low. The present spectrum has somewhat better coverage at high energies than previous hard states as observed by both \integral\/ and \xte. We notice, however, that the ISGRI and SPI data do not agree well. SPI detects significantly higher flux values above 100 keV and the slope is flatter than for ISGRI in the whole energy range. The slope difference can be due to a difference in calibration, but that does not explain why SPI detects an almost flat (in $EF_{E}$ space) hard tail when ISGRI does not. Due to the difficulties in background subtraction in the crowded Cygnus field, the SPI tail has to be taken with caution. The difference between the two instruments also limits the goodness of our fits above 100 keV, as it is impossible to find a model that would fit both data sets. 

\subsection{Physical model for the intrinsic emission}
\subsubsection{Comptonization by a hybrid electron distribution}
We use a Comptonization model, {\tt eqpair} (Coppi 1992, 1999, Gierli\'nski et al. 1999). The model assumes a physical scenario where cool seed photons from an accretion disc are upscattered via inverse Compton scattering by electrons in a hot plasma, located either inside the inner radius of, or on top of, an accretion disc. The accretion disc is assumed to emit as a multicolour blackbody ({\tt diskbb}, Mitsuda et al. 1984) with the maximum temperature $T_{\rm s}$ as a free parameter. 

The main fitting parameter in the model, the one that determines the slope of the Comptonized spectrum, is the ratio between the luminosities corresponding to the plasma heating rate, $L_{\rm h}$, and to the irradiating seed photons, $L_{\rm s}$, here expressed as their respective compactnesses, $\lh$ and $\ls$, defined as $\ell \equiv L\sigma_{\rm T}/(r m_{\rm e} c^3)$, where $r$ is the characteristic size of the X-ray emitting region, $\sigma_{\rm T}$ the Thomson cross section and $m_{\rm e}$ the electron mass. A value of $\lh/\ls \geq 1$ describes a hard state,  where plasma heating dominates over photon cooling, and $\lh/\ls \leq$1, a soft state, where the luminosity in the irradiating seed photons dominates and cooling becomes efficient. The value of the soft compactness, $\ls$, is a fit parameter as well. However, as our data, covering energies down to 3 keV only, cannot uniquely constrain $L_{\rm s}$ and thus $\ls$, we assume a constant $\ls=100$, corresponding to a luminous source with a comparably small radius (the same as used for GRS 1915+105 in Zdziarski et al. 2005 and Hannikainen et al. 2005). 

In {\tt eqpair}, the electron distribution can be purely thermal or hybrid, i.e., Maxwellian at low energies 
and non-thermal at high energies, if an acceleration process is present. This distribution, including the electron temperature, $T_{\rm e}$, is calculated self-consistently from the assumed form of the acceleration (if present) and from the energy balance between the hot electrons and the injected seed photons. For a hybrid distribution, the power supplied to the electrons in the plasma is the sum of direct heating and acceleration, so for the hard compactness, $\lh=\lth+\lnth$, where $\lth$ and $\lnth$ denote the corresponding compactnesses of thermal heating (in addition to Coulomb energy exchange with non-thermal electrons and Compton heating) and acceleration, respectively. The acceleration is assumed to have a power-law rate with index $\Gamma_{\rm inj}$, which is a free parameter in the fit, for Lorentz factors between 1.3 and 1000. 

The total plasma optical depth, $\tau_{\rm tot}$, includes contributions from electrons formed by ionization of the atoms in the plasma, $\tau_{\rm e}$ (a free fit parameter) and from \ee pairs, $\tau_{\rm tot}-\tau_{\rm e}$, calculated self-consistently by the model. The importance of 
pairs depends both on the value of the compactness and on the shape and strength of the injection spectrum at high energies. With both a high luminosity in seed photons and a significant amount of high energy injection, creating a substantial number of photons $>$ 511 keV, the plasma is likely to be pair dominated.

\subsubsection{Compton reflection}
The code {\tt eqpair} includes Compton reflection, parametrized by an effective solid 
angle subtended by the reflector as seen from the hot plasma, $\Omega/2\pi$ (Magdziarz \& Zdziarski 1995). The value $R=\Omega/2\pi$ can also be interpreted as relative reflection compared to the intrinsic component (see further discussion in Section 3.3.3). The strength of the reflected component depends on the system inclination which is not well known in Cyg X-3. The strong orbital modulation suggests a high inclination, but not too high since there is no eclipse by the secondary. For the purpose of modelling the reflection, we use $60\degr$ (Vilhu et al. 2007). We find that our fits cannot constrain ionization of the reflecting medium and thus we assume a neutral reflector.

The iron line in Cyg X-3 is unusually strong. It may be a combination of the Fe K$\alpha$ fluorescent line, arising both from reflection in the disc and from absorption in the surrounding medium. Observations by {\it Chandra} (Paerels et al. 2000) have revealed lines from both neutral, He-like and H-like iron in the spectrum. The JEM-X instrument onboard \integral\/ cannot resolve this line complex. Therefore, we fit it as a single Gaussian line, with the centroid energy, $E_{\rm Fe}$, the width, $\sigma_{\rm Fe}$, and line photon flux, $F_{\rm Fe}$, as free parameters.

\subsubsection{Complex absorption} 
The Cyg X-3 spectrum is heavily absorbed and modelling the absorption requires the presence of both interstellar and local absorbers (V03, Hj04a, Szostek \& Zdziarski 2005). Since our data only cover energies above 3 keV, we do not attempt an exact modelling of the absorption in terms of its ionization structure, density and composition. Our aim is here to get a correct understanding of the unabsorbed spectral shape and to be able to derive unabsorbed luminosities. A full understanding of the complexity of the surrounding medium in this source requires a more detailed study, taking into account additional discrete absorption and emission features, using an instrument with higher spectral resolution (Szostek \& Zdziarski in preparation). We freeze the interstellar absorber to $N_{\rm H, gal}=1.5\times10^{22}$ cm$^{-2}$ (eg. Chu \& Bieging 1973). We find that the spectrum at low energies, in all three scenarios below, requires more than one simple local absorber. We model the local absorption as a two-fold absorber with one absorbing medium with column density $N_{\rm e,0}$ fully covering the source, and another with the column density $N_{\rm e,1}$, covering a fraction $f_{1}$ of the source. Since the donor is a WR-star, the abundances in the Cyg~X-3 system are likely to differ from solar abundances. In particular, little or no hydrogen is present (van Keerkwijk et al. 1992). We therefore give the column densities as $N_{\rm e}$ rather than $N_{\rm H}$. We find, however, that altering the table abundances in {\tt xspec} does not improve or significantly alter our fits. We therefore simply use the abundances of Anders \& Ebihara (1982), but with the metal, $A_{>\rm He}$, and iron, $A_{\rm Fe}$, abundances as free parameters. The true structure of the absorber is likely to be much more complex. We do include an absorption edge with the energy $E_{\rm edge}$, and depth $\tau_{\rm edge}$, which we have found to significantly improve the fit. We interpret it as that arising from H-like iron in a surrounding ionized medium at $E_{\rm edge} = 9.28$ keV (Reilman \& Manson 1979), but allow it to vary between 9.2 and 9.4 keV due to calibration uncertainties of the JEM-X instrument. 

To account for additional attenuation caused by presumably electron scattering in the stellar wind, responsible for the strong orbital modulation, we include a Compton absorber ({\tt cabs}). We first model the data with its optical depth frozen to zero. Since our data is phase averaged, we then correct for the effect of orbital modulation, which affects the normalization by a maximum of $\sim50$ per cent at phase 0 in the hard state, by increasing the normalization by 1/0.75 times and then fit the corresponding value of the Compton column density, $N_{\rm e,scatt}$. The corrected spectra and unabsorbed luminosities then roughly correspond to phase maximum. It should be noted that all quoted luminosities in this paper are lower limits, since it is likely that the scattering optical depth is non-negligible even outside the orbit of the compact object, thus affecting the maximum phase as well.

\subsection{Models and Results}
With the large uncertainties in the parameters of the absorber, blackbody temperature etc. arising from both the complexity of the source and the limitations of the instruments, detailed physical models with many free parameters, like this one, have to be used with caution. Spectral fits suffer from many local minima, some resulting in unphysical combinations of the parameters, and the resulting minima found are strongly dependent on the input parameters. A good fit can in fact be obtained for different combinations of parameters representing very different conceptual scenarios. We find that, using the same {\tt xspec} model with different input parameters, we can fit the data in at least three different ways, all giving acceptable fits to the data, but each describing a very different physical and geometrical scenario. In the three models discussed in the next section, the {\tt xspec} model is thus the same: {\tt constant*wabs*cabs(absnd*edge(eqpair)+gaussian)}, with only the parameter values differing between the three cases. In the following, when we refer to different models, what we mean by that is thus different combinations of the model parameters, all with the same components, but representing different physical situations. Since the three models presented represent statistically three local minima, no errors are given for the parameter values. The spectral modelling in this paper does not pretend to give exact numerical values of all the parameters, but is rather as a comparison of three conceptually different ways of interpreting the true nature of the hard state of Cyg~X-3.

\begin{figure}
\centering
\includegraphics[angle=0,width=8.5cm]{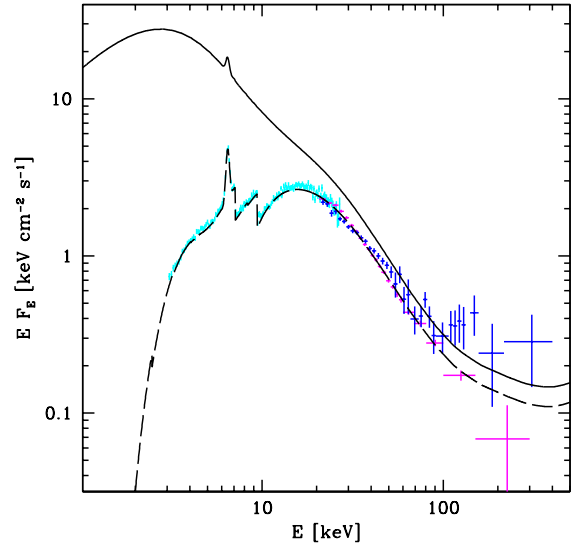} 
\caption{The data together with best-fitting model assuming strong absorption at the lower energies of an intrinsic soft spectrum. The observed, absorbed (dashes) and intrinsic, unabsorbed (solid line) model spectra are shown. Data points in cyan (3--25 keV) are from JEM-X, in magenta (20--300 keV) from ISGRI and in blue (20--200 keV) from SPI. The data from JEM-X and ISGRI were normalized to the level of the SPI normalization.}
\label{abs_data}
\end{figure}

\begin{figure}
\centering
 \includegraphics[angle=0,width=8.5cm]{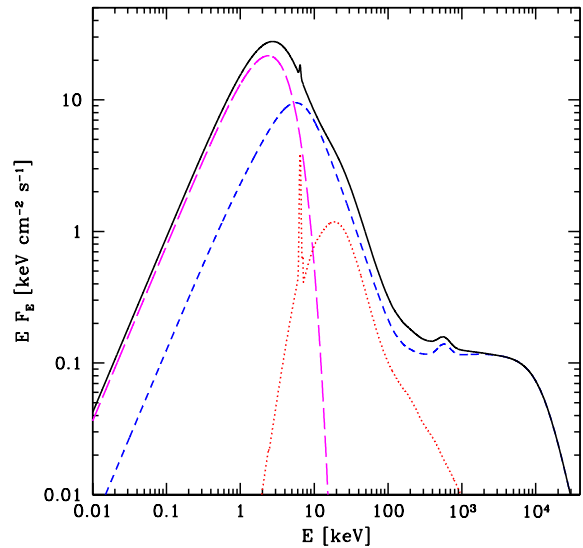} 
\caption{Components of the unabsorbed wind absorption model spectrum (solid line): the unscattered blackbody (magenta long dashes), Compton scattering (blue short dashes), Compton reflection and the Fe K-line (red dotted line).}
\label{abs_components}
\end{figure}

\subsubsection{Wind absorption model}
The assumption in this model is that the apparent hard state is not a result of a real state transition with truncation of the inner disc, but simply an effect of increased absorption of the low energy emission by local line-of-sight material in the form of a variable stellar wind. The main argument in favour of this model is the similarity of the hard state spectrum above 20 keV to that displayed in the softer states, and especially between the different spectral states as observed by \integral\/ (see Fig. \ref{groups} dashed lines and Fig. 3 in Hj04a). We know that a strong stellar wind is present in the system e.g. van Keerkwijk et al. 1992). If this wind is strongly variable, it could in principle account for the changes in the low energy end of the observed spectrum while leaving higher energies unaffected, and thus mimic a state transition.

We find that, just like in Hj04a, such a model indeed gives a good fit to the data. The model is plotted together with the data in Fig. \ref{abs_data} and the best-fitting parameters are listed in Table 1. Figure \ref{abs_components} shows the spectrum corrected for absorption decomposed into its different spectral components. The intrinsic spectrum is that of a soft state, similar to the soft states of other sources and of Cyg~X-3 itself. It is dominated by a strong blackbody component at $\sim1$ keV, presumably from an accretion disc extending all the way in close to the last stable orbit. The Comptonized spectrum is that resulting from scattering off electrons of temperature $kT_{e}=32$ keV and optical depth $\tau=0.6$, values similar to those of Cyg ~X-1 in its soft state (Gierli\'nski et al. 1999) as well as in the soft/very high state of GRS~1915+105 (Zdziarski et al. 2005). The model in addition requires strong Compton reflection and a strong iron line, likely including a strong contribution from the absorbing medium (see Section 3.2.2). It is interesting to note that despite the strong iron line, this model, as well as the reflection model in Section 3.3.3, require a low iron abundance.

The required local two-fold absorber covers 87 per cent of the X-ray source with an electron column density of $1.6\times10^{24}$ cm$^{-2}$, comparable to values observed in highly obscured AGN (Matt, Guainazzi \& Maiolino 2003), and the intrinsic spectrum is a factor of eight brighter than the observed. Depending on the exact shape of the absorption-corrected spectra of the soft states of Cyg~X-3 (Hjalmarsdotter et al. in preparation), this may indicate that, in this interpretation, the apparent lowest luminosity hard state may be the most luminous one of this source, corresponding to the highest accretion rate. Thus, not even in this interpretation of the apparent hard state is absorption alone affecting a constant underlying spectrum sufficient to explain the spectral variability. Rather, the overall spectrum would have to increase in strength at the same time as the low energies become more absorbed to explain the transition without an intrinsic pivoting or significant change of spectral shape. This is also in agreement with the observed flux and hardness anticorrelations in Section 2.1. 

To account for the (somewhat dubious) SPI points above 100 keV, this model (as well as the Reflection model in Section 3.3.3) includes a nearly flat non-thermal tail. If such a tail is real, it may be strong enough to be detected by {\it GLAST}/LAT after one year of operations (according to the {\it GLAST}/LAT performance web page \footnote{{\tt www-glast.slac.stanford.edu}}). Due to the discrepancy between the ISGRI and SPI data (see Section 3.1), we refrain from further interpretation of the strength and slope of such a tail here. There are, however, indications that such a tail may indeed be present at a similar level in at least one other state of this source (red in Fig.\ref{groups}, and Hjalmarsdotter et al. in preparation) We note that the existence of such a tail or not is, although highly interesting, not of crucial importance to our current study since it cannot by itself be used as a state indicator.

\begin{figure}
\centering
\includegraphics[angle=0,width=8.5cm]{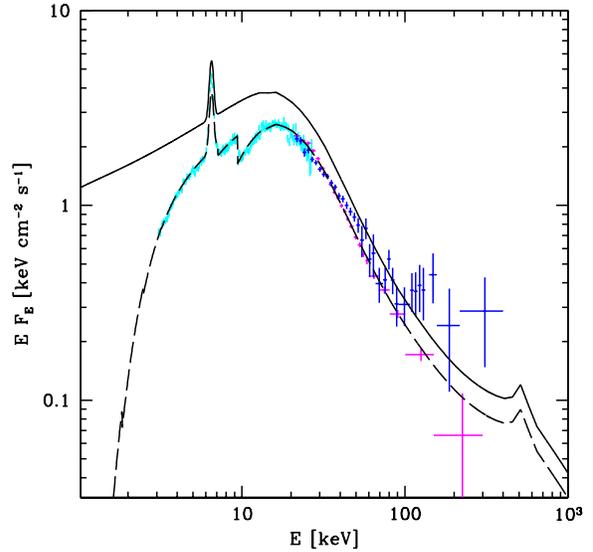} 
\caption{The data together with best fit model of a hard state with a non-thermal energy injection. The observed, absorbed (dashes) and intrinsic, unabsorbed (solid line) model spectra are shown. Data points in cyan (3--25 keV) are from JEM-X, in magenta (20--300 keV) from ISGRI and in blue (20--200 keV) from SPI. The data from JEM-X and ISGRI were normalized to SPI.}
\label{hard_data}
\end{figure}

\begin{figure}
\centering
\includegraphics[angle=0,width=8.5cm]{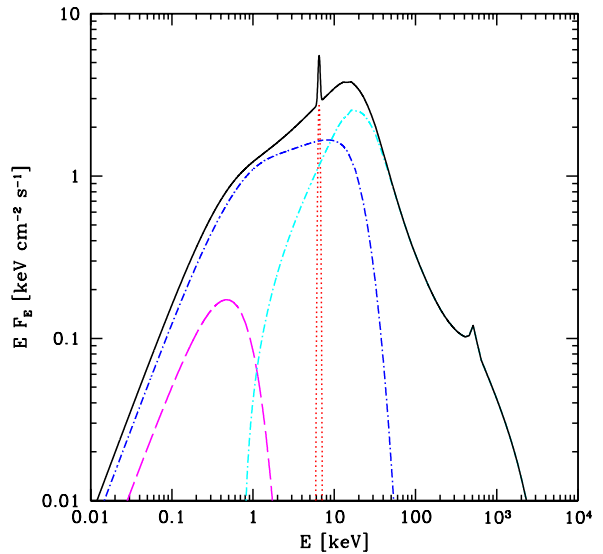} 
\caption{Components of the unabsorbed non-thermal model: the unscattered blackbody (magenta long dashes), Compton scattering from thermal (blue short dashes) and non-thermal (cyan short dashes) electrons, and the Fe K line (red dotted line).}
\label{hard_components}
\end{figure}

\subsubsection{Non-thermal model}
The main problem in identifying the hard state of Cyg~X-3 as a real hard state is the fact that its spectrum peaks at much lower energies than in other sources, implying a much lower electron temperature. Nevertheless, several results from the previous section seem to suggest a state transition similar to that of other sources. Could this still be a `real' hard state in the sense that the inner disc is truncated but without the presence of high temperature electrons? 

The universal cut-off at $\sim 50$--100 keV observed in both Galactic black holes (Zdziarski \& Gierli\'nski 2004) and in Seyfert I type AGN (Zdziarski et al. 1996), suggests that the transition into a hard state always occur at electron temperatures of this order. Neutron star binaries may have somewhat lower electron temperatures, but temperatures below 10 keV are only found in soft states (eg. Gierli\'nski \& Done 2002). Indeed, any attempt to model the spectral break at $\sim 20$ keV of the Cyg X-3 hard state with thermal Comptonization requires an electron temperature of the order of a few keV, more in accordance with temperatures observed in ultrasoft states. Our preliminary results in addition suggest that such a low temperature would indeed be lower than in the soft states of Cyg~X-3 itself, which would not be consistent with a transition to a hard state according to our present understanding.  

If we instead assume that most of the dissipated energy in the Comptonizing flow goes into particle acceleration, rather than thermal heating, the electrons will be accelerated out of a Maxwellian into a power law distribution. In {\tt eqpair}, the importance of acceleration is parametrized by the ratio, $\lnth/\lh$, between the non-thermal and total hard compactness (see Section 3.1). Given that most of these accelerated electrons will Compton cool before they thermalize, the resulting spectrum will be that of Comptonization from a hybrid distribution of electrons. Such a distribution can be described as a Maxwellian at low energies, since those electrons have sufficient time to thermalize by Coulomb interactions, plus a power law tail at high energies. The Comptonized component will thus be made up by scattering off both thermal and non-thermal electrons. In a situation where the spectrum is dominated by non-thermal Comptonization, the peak of the Comptonized spectrum is no longer determined by the maximum electron temperature but by the energy at which most electrons are injected. The shape of such a spectrum can vary significantly depending on the form of the energy distribution of the injection spectrum (mono-energetic or power law with different slopes).

We find that applying a model with almost entirely non-thermal injection in the form of a steep power law with $\Gamma=4$ and $\gamma_{\rm min}=1.3$ ($\gamma_{\rm max}$ being unimportant for such a steep injection), gives a good fit to the Cyg X-3 hard state spectrum. The model is plotted together with the data in Fig. \ref{hard_data}. The model reproduces well the observed peak at $\sim 20$ keV and the shape of the high energy slope without the addition of reflection. The equilibrium temperature of the Maxwellian electrons is only 4 keV and thus no thermal signature of high temperature electrons is present. The optical depth of the thermal part of the distribution is rather high but does not include a significant fraction of \ee pairs, despite the high compactness, and $\tau$ is dominated by the electron optical depth in the hot plasma. This does not, however, imply that \ee pairs are not produced, only that they annihilate 
at a fast rate due to the high optical depth and in fact all our models predict a strong annihilation line at 511 keV. Such a line has yet to be confirmed in any X-ray binary. With access to good coverage at high energies by OSSE, the abscence of any annihilation line in the soft state spectrum of Cyg~X-1 could, as shown by Gierli\'nski et al. (1999), limit the compactness in this source to $\la 10$. The much higher luminosity of Cyg X-3, together with a probably much smaller Comptonizing region, due to the small size of the system, is suggestive of a significantly higher value of the compactness, and the presence of an annihilation line cannot be ruled out until sufficiently good data at high energies is available. If the line is indeed not present, this would limit the compactness to a lower value in this model which would result in a somewhat worse fit.

In Fig. \ref{hard_components}, the components of the intrinsic model are plotted. To show the importance of non-thermal Comptonization, we have plotted the thermal and non-thermal contributions separately. This is done by setting the non-thermal compactness to zero, $\lnth=0$, and adjusting the cooling rate, $\lh/\ls$, to calculate the thermal spectrum that would have the same $\tau$ and $kT_{e}$ as the original hybrid model and then subtract it from the total model, in the same way as explained in Hannikainen et al. (2005).

With the \integral\/ data, covering energies above 3 keV only, the disc seed photon temperature, especially in this model assuming a low temperature, is at best badly constrained and very much dependent on the assumed strength and shape of absorption, leaving the parameters of the absorber and $T_{\rm s}$ to a large extent degenerate. We find, however, that a good fit can be achieved with a disc temperature of 200 eV, equal to that of Cyg X-1 in its hard state. Such a low disc temperature would suggest that the disc is truncated at a radius $r \gg 6r_{g}$, the last stable orbit.

Despite the absence of any continuum reflection feature, the presence of a strong iron line is required by the data. In this model it is thus likely that the iron line arises from flourescence in the absorbing medium. 

\begin{figure}
\centering
\includegraphics[angle=0,width=8.5cm]{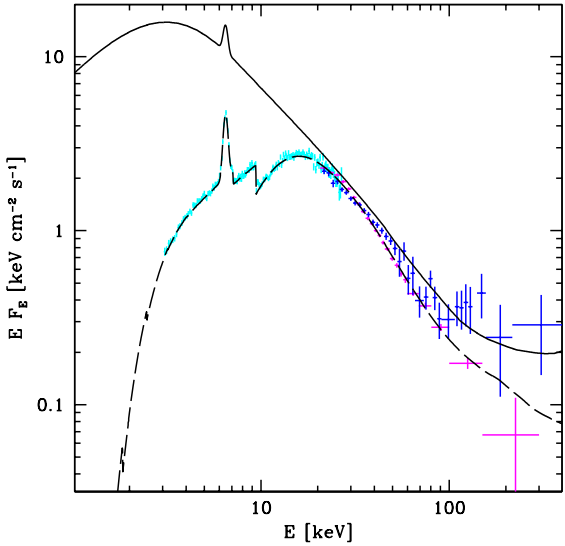} 
\caption{The data together with best-fitting model describing an almost pure reflection spectrum (dashes) of an intrinsically soft spectrum (solid line). Data points in cyan (3--25 keV) are from JEM-X, in magenta (20--300 keV) from ISGRI and in blue (20--200 keV) from SPI. The data from JEM-X and ISGRI were normalized to SPI.}
\label{refl_data}
\end{figure}

\begin{figure}
\centering
\includegraphics[angle=0,width=8.5cm]{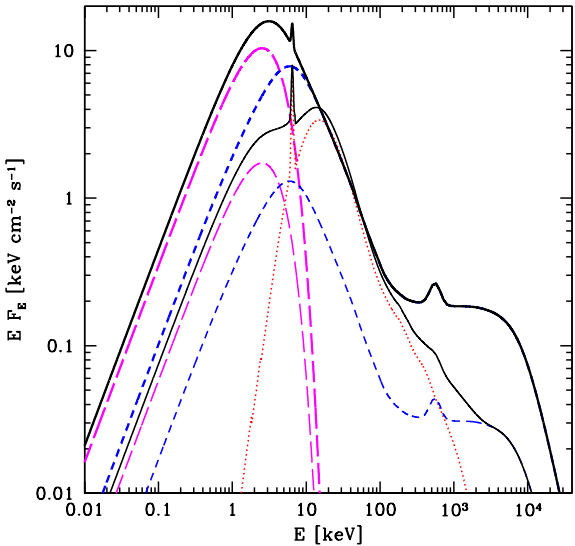} 
\caption{The observed spectrum, corrected for absorption, is made up by the part of the unscattered blackbody (magenta long dashes) and its Comptonization (blue short dashes) seen directly, plus the strong Compton reflection component (red dotted line). The much stronger required incident spectrum before reflection is shown by the heavy solid line with heavy magenta long dashes for the unscattered blackbody and heavy blue short dashes for the Comptonized part.}
\label{refl_components}
\end{figure}

\subsubsection{Reflection model}
We note that the shape of the Cyg X-3 spectrum in its apparent hard state is in fact very much reminiscent of that arising from Compton reflection off a cool medium. The signatures of Compton reflection, an Fe K$\alpha$ line, an Fe absorption edge and a pronounced hump in the spectrum at energies above 10 keV, seem to be required in fits of most spectra of both X-ray binaries and AGN. Compton reflection arises when the energetic upscattered photons from the hot plasma scatters off the cooler accretion disc and becomes more important with a larger solid angle covered by the disc as seen by the X-ray source (Gilfanov, Churazov \& Revnivtsev 1999; Zdziarski, Lubi\'nski \& Smith 1999). In some cases, the central source can be entirely obscured and the observed spectrum that of reflection alone. This is the case for the so-called obscured AGN, active galaxies of type Seyfert-2, hidden behind a Compton thick absorber (e.g. Matt et al. 1996). A Compton reflection dominated spectrum has also recently been suggested to explain a low non pulsating state of the neutron star GX~1+4 by Rea et al. (2005). The proposed geometry is a thickened quasi-toroidal disc, hiding the central source from view but giving rise to strong reflection. In such a situation, $R$ can largely exceed $2.0$ and is then rather a measure of how much more of the emission that is seen only after reflection compared to what can be observed directly, than a measure of a physical solid angle.

We find that while a pure reflection spectrum does not well describe the lowest energy data of the Cyg X-3 spectrum, which show significant excess below 10 keV, a model dominated by reflection, with the addition of only a weak intrinsic component, reproduces the observed spectrum very well. The model is plotted together with the data points in Fig.~\ref{refl_data} and corresponds to a situation where $R=12$. Figure~\ref{refl_components} shows the spectral components, including the components of the required incident spectrum before reflection as calculated by {\tt eqpair}. We see that for the reflected spectrum to match the shape of the observed spectrum of Cyg~X-3, the incident spectrum has to be a soft one, slightly harder than the intrinsic spectrum of the absorption model, but probably very similar to one of the observed soft states of Cyg~X-3 (cf spectrum in magenta in Fig.~\ref{groups}), once corrected for absorption (Hjalmarsdotter et al. in prep.). To give rise to such a strong reflection component, the incident spectrum also has to be very luminous, and most likely more luminous than the soft states of the source (again by mere comparison with the flux levels of the soft state spectra in Fig.~\ref{groups}). The exact strength depends on the multiplication factor used, which for $R=12$ can be anything $\geq 6$, depending on the exact geometry. We have used 6, corresponding to a large covering factor of the reflector as seen by the X-ray source, and thus giving a minimum possible luminosity for the incident spectrum, that may be even higher. Since the thickening of the disc is likely to be a response to very high, possibly super-Eddington, accretion rates (Shakura \& Sunyaev 1973, Jaroszy\'nski et al. 1980), a stronger incident spectrum in the 'reflected' state is expected in this scenario. This is also in agreement with the observed anticorrelations of hard and soft X-ray fluxes which, just as in the absorption model, require the whole spectrum to become stronger to match the increase of hard X-rays if the shape is not changing in the transition. In this interpretation, a large portion of the spectral variability observed in this source could be modelled by varying the relative amount of emission seen only in reflection against that which can be observed directly, with an underlying intrinsic spectrum varying only in normalization. 

We note that this model gives the best formal fit to the present data, followed by the absorption model and the non-thermal model in order of increased $\chi^{2}$. Due to the limited energy range and our approximate treatment of absorption, we however do not believe that the presented models can be ranked based on formal statistics, but should be evaluated physically together with our other results concerning the behaviour of the source. In the next section, the implications of the models and their agreement with our other results will be discussed.

\begin{table*}

\caption{Model parameters. Since the models represent three local minima, no errors are given.}
\begin{tabular}{lcccccc}
\hline
 & 				  	&					{\rm Wind absorption model} &	{\rm Non-thermal model} &		{\rm Reflection model}\\
 \hline
$N_{\rm e,0}$& 	$10^{22}\, {\rm cm}^{-2}$ &	$11$ &					$3.2$ &						$6.5$\\ 
$N_{\rm e,1}$& 	$10^{22}\, {\rm cm}^{-2}$ &	$150$ &					$3.0$ &						$110$\\ 
$f_{1}$ &			 &						$0.87$ &					$0.16$ & 						$0.16$ \\
$N_{\rm e,scatt}$ &	$10^{22}\, {\rm cm}^{-2}$ & 	$42$ &					$42$ &						$42$ \\
${\rm A_{Fe}}$ & 	 &						$0.58$ &					$3.20$ &						$0.35$\\
${\rm A_{>He}}$ &	&						$3.6$&						$1.29$&					$1.23$\\
$E_{\rm edge}$ & 		${\rm keV}$&				$9.4{\rm a}$&				$9.4{\rm a}$ &  	 				$9.4{\rm a}$\\
$\tau_{\rm edge}$ & 	 &						$0.50$ &					$0.34$ &						$0.39$\\
$kT_{\rm s}$ & 		${\rm eV}$&				$1014$ &					$200$&						$1069$\\
$\lh/\ls$  &  		 &						$0.26$&					$7.14$&						$0.44$\\
$\lnth/\lh$ & 		 &						$0.07$&					$0.96$ &						$0.12$\\
$\Gamma_{\rm inj}$ &	&						$2.0$&					$4.0$&						$2.0$\\	
$\tau_{\mathrm e}$ & 		 &						$0.58$&					$9.43$ &						$1.00$\\
$\tau_{\mathrm tot}^{\mathrm{b}}$ & 		 &						$0.61$&					$9.50$ &						$1.09$\\
$kT_{\mathrm e}^{\mathrm{b}}$ & ${\rm keV}$& 		32 &						4.0 &							24\\
$R=\Omega/ 2\pi$ & 	 &						$1.3$ &					$0.0$ &						$12$\\
$E_{\mathrm Fe}$ &			${\rm keV}$& 				$6.46$ &					$6.49$ &						$6.48$\\
$\sigma_{\mathrm Fe}$ & 			 &						$0.14$ &					$0.18$ &						$0.18$\\
$F_{\mathrm Fe}$	& ${\rm cm}^{-2}\, {\rm s}^{-1}$	&	$2.8\times 10^{-2}$	&	$2.9\times 10^{-2}$	& $4.0\times 10^{-2}$\\ 
$F_{\rm bol}^{\rm c}$ & 	${\rm erg\, cm}^{-2}\, {\rm s}^{-1}$ & $8.2\times 10^{-8}$ &		$1.9\times 10^{-8}$ &			$6.6\times 10^{-8}$\\
$L_{\rm bol}^{\rm d}$& ${\rm erg}\, {\rm s}^{-1}$& 	$7.9\times10^{38}$ &		$1.8\times10^{38}$ &			$6.4\times10^{38}$\\
$\chi^2/\nu$		 &					&	269/207 &					281/207 &						259/207\\
\hline
\end{tabular}
\begin{list}{}{}
\item[$^{\mathrm{a}}$] Allowed between 9.2 and 9.4 due to calibration uncertainties of the JEM-X instrument. 
\item[$^{\mathrm{b}}$] Calculated from the energy balance, i.e., not a free fit parameter. 
\item[$^{\mathrm{c}}$] The bolometric flux of the unabsorbed model spectrum normalized to the SPI data.
\item[$^{\mathrm{d}}$] The unabsorbed bolometric luminosity assuming a distance of 9 kpc.
\end{list}
\end{table*}

\section{Discussion}

\subsection{State transitions vs absorption}
The spectral variability in X-ray transients and in the best-studied Galactic black hole Cyg X-1 is generally believed to be driven by transitions between two stable configurations of the accretion flow: a high accretion rate disc-dominated soft state and a low accretion rate, possibly advection-dominated, hard state, where the inner disc is truncated and replaced by a hot flow. Many X-ray transients show a wider range of spectral variability during outbursts, including several intermediate states and soft states with varying amplitude of the hard tail (see e.g. examples in Done \& Gierli\'nski 2003). The changes between soft disc-dominated and hard non disc-dominated states that are also accompanied by a distinct change in radio emission and jet properties (e.g. Fender, Belloni \& Gallo 2004 and references therein).

The spectral variability of Cyg X-3 is similar to some transients (see e.g. a comparison with the spectra of XTE J1550--564 in Zdziarski \& Gierli\'nski 2004) and shows a distinct bimodality between soft and hard states. In Cyg~X-3 the observed spectral states cannot, however, be unambiguously identified to those of other sources due to presence of strong absorption largely affecting the observed spectra and possibly strong enough in some states to mimic state transitions. Unlike the transients, but like e.g. Cyg~X-1, Cyg~X-3 is a wind-fed system. The difference between Cyg~X-3 and Cyg~X-1 as well as other wind-fed systems, is the extreme proximity to its companion and the presence of a strong wind. The Cyg~X-3 system is very compact with the compact object located almost inside the extended photosphere of its companion. The situation differs from normal wind accretion in that there is indeed much more matter available to accrete. With such a constant high supply, it is possible to imagine that the source would always be in a high accretion rate and not experience transitions to some lower accretion rate hard state. Nevertheless the results from Section 2 show several signs thereof.

The bimodal flux distribution in Cyg~X-3 reflects the existence of two stable configurations, suggestive of a bimodal behaviour of the accretion disc. In a wind accreting system, state transitions are indeed believed to be a response to a change in wind parameters and mass loss of the companion causing a re-configuration of the accretion flow at some limiting accretion rate. But for the bimodality to be caused by absorption effects, the wind has to be strictly bimodal by itself and switch between an `on' and an `off' state with strong mass-loss and high absorption to cause the apparent hard state and a lower mass-loss and less absorption in the soft state. 

The existence of the anticorrelations between hard and soft X-rays and between flux and hardness further show that the spectral variability cannot be due to simple variability of absorption with an underlying intrinsic spectrum of constant shape and strength. For an increase of hard X-rays in the hard state, the intrinsic spectrum and the accretion rate has to be higher in the hard state in the wind absorption scenario (as well as in the reflection model). 
An intrinsic pivot somewhere between 12 and 20 keV would instead naturally explain the observed anticorrelations without any variable absorption.

The fact that the shape of the orbital modulation does not change significantly between the states and that the modulation becomes {\it stronger} in the soft states also speaks against any increase in optical depth or a re-distribution of the local absorbing matter in the transition to the hard state. If anything, both this study and preliminary results based on {\it Chandra} data (McCollough et al. in prep.) suggest that the wind is stronger and mass-loss larger in the soft states.

Finally, the existence of a very similar pattern of the radio/X-ray correlation in the Cyg~X-3 hard state to that of other sources, with the sharp change between radio behaviour at exactly the same flux level as that corresponding to its apparent transition, is probably the strongest argument for a real state transition in Cyg~X-3. Since the radio emission originates from jets far away from the compact object and several orders of magnitude further out than the size of the system, these effects cannot be due to absorption.

The reflection model, proposed in Section 3.3.3, offers an alternative `in between' the absorption model and a real transition. It involves a real change in the accretion flow, with the thickening of the disc, possibly triggered by super-Eddington accretion rates, which may well show the observed bi-modality. A spectral variability explained by varying the fraction of the emission seen direct and only in reflection, respectively, would explain the observed anticorrelations between flux and hardness. Since the change in the accretion flow is intrinsic with respect to the surrounding matter, the model is also in agreement with the orbital modulation not increasing in the hard state. The reflection model however suffers from the same problem as the absorption model in that the required incident spectrum is very bright and soft, with a strong disc component present. This implies a higher accretion rate in the apparent hard state and does not seem to agree with for example the radio/X-ray correlations, even if the change in radio behaviour in other systems could, in principle, be a response to other properties of the hard state than the disappearance of the inner disc. 

We note that an additional, independent way to investigate the existence of a state transition and truncation of the inner disc, would be a study of the timing behaviour. Such a study could not be done using the data presented here due to its limited time resolution, but will be presented elsewhere, using other data (Hjalmarsdotter et al. in prep.).

\subsection{The origin of the non-thermal electrons}
While it is true that the observed spectra of hard states of Galactic as well as extragalactic black hole sources are usually well described by purely thermal Comptonization, in soft states the non-thermal fraction is often significant or dominating. Extended flat tails in both eg. Cyg~X-1 (McConnell et al. 2002) and GRS~1915+105 (Zdziarski et al. 2001), with no apparent cut-off below 1 MeV, serve as strong evidence for the presence of non-thermal electrons accelerated to supra-thermal energies in these sources. The precise acceleration mechanism responsible for the observed non-thermal emission observed in soft states of Galactic X-ray binaries as well as in AGN is not well known. A part of the problem is our current lack of understanding of the soft state beyond the thin disc approximation, in particular the location and physical conditions of regions responsible for hard X-ray production in that state.

The disc in Cyg~X-3 is located in a highly turbulent environment. It may be subjected to constant collisions with the wind even at small radii, close to the compact object. Such collisions are likely to produce shocks resulting in particle acceleration in both the thin disc and perhaps also in the inner advective flow of a hard state. In fact, it can be shown that some observations of Cyg X-1 in its hard state seem to require some small fraction of non-thermal electrons, in the form of weaker component in addition to the main thermal one (e.g. McConnell et al. 2002, Ibragimov et al. 2005). Ibragimov et al. (2005) found that the addition of such an additional non-thermal component could explain not only the high-energy tail as observed by \osse\/, but in addition the soft excess of unknown origin required in fits to the hard state of Cyg~X-1 and usually modelled with an additional thermal Comptonization component. With the improvement of gamma-ray instruments, more such hard tails are likely to be found. 

In the scenario of the non-thermal model, the peculiarity of the Cyg X-3 hard state is thus perhaps not the presence of non-thermal emission, but rather the absence of a strong hot thermal component. This means that whatever acceleration mechanism is at work, it must be efficient enough to dominate entirely over and prevent thermal heating of the plasma electrons. To understand how this works, the effect of a strong wind on an advective flow should be studied in detail.  

If the 'hot' inner flow in a truncated disc geometry is in essence the same region as the base of the radio emitting jets, a highly efficient acceleration process may also be responsible for the unusually strong radio emission of Cyg~X-3. If it is also occasionally capable of accelerating electrons up to very high energies, it could explain several early but unconfirmed claims of Cyg~X-3 as a source of ultra-high energy gamma-rays (see the review by Bonnet-Bidaud \& Chardin 1988 and references therein). 

\subsection{Luminosities and implications for the mass of the compact object}

The Eddington luminosity for the X-ray emitting compact object is: 
\begin{equation}
L_{\rm E}\equiv 4\pi \mu_e GM m_p c/\sigma_{\rm T},
\end{equation}
 where $\mu_e=2/(1+X)$ is the mean electron molecular weight, $X$ the hydrogen mass fraction, and $\sigma_{\rm T}$ the Thomson cross section. With the companion being a Wolf-Rayet star, the accreted matter contains no or very little hydrogen, and $\ledd=2.5\times 10^{38} M/\msun$ erg s$^{-1}$. 

Table 1 shows that both the absorption and the reflection models give a bolometric X-ray luminosity exceeding the Eddington luminosity for a $1.4 \msun$ neutron star (assuming a distance of 9 kpc). 
Both luminosities are within the range for typical luminosities of a stable soft state for a $\sim10\msun$ black hole and consistent with the soft spectral shapes. The high luminosities of these models are thus not a problem in themselves, even if the fact that these models imply higher luminosities for the apparent hard than for the observed soft states of the source does not seem to be in agreement with our other results. In these two models, a neutron star accretor cannot be ruled out if accretion is super-Eddington, as explicitly suggested by the reflection model. For example the neutron star X-ray binary Cir~X-1 can be as luminous as $\sim 10 L/\ledd$ in its soft state (Done \& Gierli\'nski 2003 and references therein).

Without the presence of hydrogen in the accreted matter, the non-thermal model gives a luminosity below $\ledd$ for a $1.4 \msun$ neutron star. This model is however not consistent with a neutron star accretor since it describes a low accretion rate hard state, which is observed in atoll systems only at luminosities below $\sim 10$ per cent of $\ledd$ if no hysteresis effects are present (Gladstone, Done \& Gierli\'nski 2007). As a persistent system, Cyg~X-3, just like Cyg~X-1, does not show any hysteresis and the luminosity for the state transition is constant and the same for the hard-to-soft as for the soft-to-hard transitions. If the non-thermal model is a correct interpretation of the Cyg~X-3 hard state, a neutron star accretor can thus be ruled out. For black holes, both observations (eg. Done \& Gierli\'nski 2003) and theoretical models of advective flows (Esin, McClintock \& Narayan 1997) suggest that the transition to a hard state takes place at a maximum of a few percent ($\sim 3 \%$) of the Eddington luminosity (again if no hysteresis effects are present). In order for the luminosity in this model to be a small enough fraction of $\ledd$, the mass of the compact object would have to exceed $20\msun$. This could make Cyg~X-3 the most massive black hole observed in an X-ray binary in our Galaxy so far. If this non-thermal hard state is allowed at slightly higher luminosities, the mass could be consistent with the estimate of $17 \msun$, by Schmutz et al. (1996).

Luminous hot accretion flows (LHAF) have been investigated by Yuan et al. (2007) and shown to be a able to explain the large luminosities in excess of 10 per cent of the Eddington luminosity seen in the hard state of some transient black hole binaries. The properties of a non-thermal advective flow have not yet been studied. In addition, the complex environment in the Cyg~X-3 system compared to other systems may make disc formation difficult and the conditions for stable disc accretion may depend not only on the properties of the accretion solution but also be limited by dynamical constrains, making it possible to support a stable inner disc only at higher $L/\ledd$ than in other systems.

\section{Conclusions}
We have studied the X-ray binary Cyg~X-3 in its apparent hard state, characterised by low ASM flux levels (1--12 keV) and a spectrum peaking at $\sim20$ keV. Our aim was to determine whether Cyg~X-3 experiences a state transition from a disc-dominated state to a state where the inner disc is truncated and replaced with a hot flow, or whether this apparent transition is just an effect of increased local absorption. 

From a study of the ASM light curve we find that the distribution of soft X-ray fluxes is bimodal, reflecting the constant switching between this apparent hard and a soft state, with a variable amplitude of the high energy tail. The distribution within each peak is log-normal. The previously reported anticorrelations between hard and soft X-rays and between soft X-ray flux and hardness are present in both states and are indicative of an intrinsic spectral pivoting rather than of variable absorption. The radio/X-ray correlations in Cyg~X-3 are very similar to that of other sources with a distinct change in the radio behaviour at the flux level marking the state transition. A study of the orbital modulation further shows that an increase in wind optical depth in the hard state is unlikely. These results are all strongly suggestive of a real state transition in Cyg~X-3, with a change in the configuration of the accretion flow, rather than variability of absorbing local wind material in the line of sight. 

We model the broad-band X-ray spectrum in the apparent hard state, using \integral\/ data. Due to the strong absorption indeed present in this system, spectral modelling is not straightforward or unique, and several interpretations are possible. The main problem in identifying the apparent hard state in Cyg~X-3 with a `real' hard state has been the unusually low high-energy cut-off, implying a very low electron temperature, not consistent with a transition to a `hot flow'. We find that allowing for the accreted power to be transferred to the electrons in the plasma in the form of acceleration rather than thermal heating, thus producing a spectrum dominated by non-thermal Comptonization, usually observed in soft states, gives a good description of the Cyg~X-3 hard state spectrum. The input seed photons originate in a weak low temperature blackbody from an accretion disc truncated far away from the innermost stable orbit. The luminosity in this non-thermal hard state is as high as $\sim$ 10 per cent of the Eddington luminosity for a $10 \msun$ black hole. This means that either it is much more radiatively efficient than the standard ADAF models, or the compact object is very massive, $> 20 \msun$.
  
We find that a wind absorption model, where the apparent hard state is created by increased absorption at low energies, as well as a reflection model, where the spectrum of the hard state is interpreted as that of almost pure Compton reflection in a geometry with a thickened disc, obscuring the direct emission from view, also both give a good fit to the data, both formally better than our preferred model. However both of these models imply a very luminous and soft intrinsic spectrum, not in agreement with our other results.

In conclusion, we suggest that the observed hard state in Cyg~X-3 is indeed a hard state with truncation of the inner disc and involves a change of accretion solution and disc geometry. The absence of hot electrons giving rise to a thermal Comptonization peak at $\sim$ 100 keV, as seen in most other sources in their hard states, is explained by the non-thermal nature of this hard state, with the accreted power being supplied to the plasma electrons in the form of acceleration rather than thermal heating. This unusual behaviour may be an effect of the strong wind interacting with the accretion flow in this extremely compact system.

\section*{Acknowledgements}
We thank J. Kn\"odlseder, G. Rauw and I. Stevens for their kind permission to use the Cyg X-3 \integral\/ data from revolutions 251, 252 and 255, and G. Pooley for providing the Cyg X-3 radio data from the Ryle telescope. We thank A. Paizis for reducing the IBIS/ISGRI data. 
LH thanks P. Hakala for creating a folding tool for the evolving Cyg X-3 period, M. Axelsson and L. Borgonovo for help with Fig. \ref{hist} and for stimulating discussions, as well as A. Szostek for sharing of thoughts and results on Cyg X-3 in general. LH and DH acknowledge support from the Finnish Academy, and LH in addition from the V\"ais\"al\"a foundation. AAZ and LH have been supported in part by the Polish grants 1P03D01827, 1P03D01128 and 4T12E04727. SL is grateful for support from the Swedish National Spaceboard. MM was supported in part by NASA contract NAS8-03060. This work was partly funded by NORDITA through the Nordic Project in High Energy Astrophysics. Based on observations with \integral, an ESA project with instruments and science data centre funded by ESA and member states (especially the PI countries: Denmark, France, Germany, Italy, Switzerland, and Spain), the Czech Republic, and Poland and with the participation of Russia and the US. Finally, we acknowledge quick-look results provided by the \xte\/ team.

\label{lastpage}

\end{document}